\documentclass[12pt]{article}
\newcommand{\nnn}{\noindent}
\newcommand{\oo}{\over}

\newcommand{\be}{\begin{equation}}
\newcommand{\ee}{\end{equation}}

\newcommand{\vs}{\vspace}
\newcommand{\hs}{\hspace}

\newcommand{\pp}{{\bf p}}
\newcommand{\ssb}{{\bf s}}
\newcommand{\rr}{{\bf r}}

\def\bear{\begin{eqnarray}}
\def\ear{\end{eqnarray}}

\begin{document}
\baselineskip .5cm

\pagestyle{myheadings}
\markboth{International Journal of Theoretical Physics, 
Vol.\,9,\,No.\,4\,(1974),\,pp. 229--244}
{International Journal of Theoretical Physics, 
Vol.\,9,\,No.\,4\,(1974),\,pp. 229--244}
 
\ 

\vs{1.5cm}

\begin{center}

{\bf \LARGE External Inversion, Internal Inversion, and Reflection Invariance}

\vs{6mm}

Matej Pav\v si\v c

Institute J. Stefan, University of Ljubljana, Ljubljana, Yugoslavia

{\it Received: 18 June 1973}

\vs{1.5cm}

{\it Abstract}

\end{center}

Having in mind that physical systems have different levels of 
structure we develop the concept of external, internal and total 
improper Lorentz transformation (space inversion and time reversal). 
A particle obtained from the ordinary one by the application of 
internal space inversion or time reversal is generally a different
particle. From this point of view the intrinsic parity of a nuclear
particle (`elementary particle') is in fact the external intrinsic
parity, if we take into account the internal structure of a particle.
We show that non-conservation of the external pa\-rity does not
necessarily imply non-invariance of nature under space inversion.
The conventional theory of beta-decay can be corrected by
including the internal degrees of freedom to become invariant
under total space inversion, though not under the external one.

\vs{6mm}

{\bf Note}: {\it This old paper has recently been subjected to
considerable interest due to the increasing exploration
of the ``exact parity model", ``mirror matter" and
its possible manifestation in astrophysics. It is not
available elsewhere on the Internet, and in the interests
of easy access I am posting it on {\tt astro-ph}. The body
of the paper is identical to the published version,
except for few minor correction of spelling.}

\vs{6mm}

\nnn Preseent address: J. Stefan Institute, Jamova 39, SI-1000
Ljubljana, Slovenia

\nnn E-mail: matej.pavsic@ijs.si

\newpage

\section{Introduction}

Though the concept of space inversion is clear to us from the geometrical
point of view, we must be careful when applying space inversion to
real physical objects. Classical examples show that classical
particles have an internal structure which must also be transformed
under space inversion. If only positions, translatory and angular
momenta are inverted, the transformation is not a complete space
inversion, but only a partial one. In the domain of nuclear particles
(`elementary particles') we have become accustomed to consider
space--time coordinates of particles as one thing and the eventual
particle's structure in its internal space as the other thing, independent
of space--time.We have believed that when reversing positions,
momenta and angular momenta of particles, we have achieved
complete space space inversion. Experiments show that the
proton has an internal electromagnetic structure. Particles do not
differ among themselves only in their space--time properties
(spins for instance) but also in other properties, which indicate their
internal structure. There is no reason why we should not admit
that this internal structure is also due to space--time transformations.
We develop the concept of external, internal and total improper
Lorentz transformation. We then postulate that each physical
theory must be invariant under total improper Lorentz transformation,
though not necessarily under an external or internal one.

We apply these ideas to beta-decay. As is well known, the
distribution of electrons emitted by oriented Co$^{60}$ nuclei
at the beta-decay is asymmetric with respect to the axis of
orientation (Fraunfelder {\it at al.}, 1957). Polarised electrons
are emitted preferentially in one direction and anti-neutrinos in
the opposite direction. For electrons and anti-neutrinos have
well-defined helicities, the total system has a definite handedness.
The mirror picture of the beta-decay is different from the original
picture. No such mirror decay was observed, and it was concluded
that parity is not conserved in the beta-decay. Hence the weak
interaction must contain pseudoscalar terms.

Nature, however, has always appeared to be symmetric in its basic
laws, but we have suddenly an unpleasant asymmetry with respect
to space inversion. Several attempts have been made in order to
save the invariance. One of the proposals was (Salam, 1957) to
suggest that every particle has its double which differs from it
by its `handedness'. But, unfortunately, no distinction is made
between external and internal space inversion (see also the following
text), therefore the theory fails to be convincing for other particles
than neutrinos. Our point of view is different from that of Salam.
The purpose of this paper is to show how the apparent asymmetrical
behavior of the weak processes with
respect to space inversion could be explained as the symmetrical
behavior by having in mind three types of space inversion:
external, internal and the total one.

The symmetric behavior is automatically obtained if we postulate
the new kind of particles that are obtained from the ordinary
particles by applying to the latter the internal inversion $P_I$.
Let $a$ (or $a_+$) be an ordinary particle, say proton or neutron,
etc., and $a_-$ a particle obtained by internal space inversion
(a mirror particle). Two particles are related by the internal
space inversion $P_I$ in the following way
$$a_+ \rightarrow a_- = P_I a_+ \; , \qquad P_I a_- = a_+$$
If a particle $a_+$ has a definite helicity $({\bf s}\cdot {\bf p})$,
a particle $a_-$ has the same helicity, but differs from $a_+$
in its internal structure. What we would like to say is that it is
not correct to assume that a mirror image of an elementary particle
is generally the same and behaves in the same way in reactions as
a particle. But this {\it a priory} assumption has been achieved
whilst interpreting the asymmetric Co$^{60}$ beta-decay as
the proof for the mirror asymmetry of the weak interaction.
By saying that the mirror beta-decay is an impossible process, we
tacitly assume that protons or neutrons in the mirror are the same
protons or neutrons. There is no experimental evidence for such
an assumption. On the contrary, the existence of the anomalous
proton or neutron magnetic moments indicates the asymmetric internal
structure of two particles. Hence it is possible that the mirror
beta-decay exists, but protons or neutrons that decay are mirror
protons or mirror neutrons. Therefore, instead of saying that
the mirror picture of the processes

\

\hfill
$\begin{array}{l}
          p \rightarrow n + e^+ + \nu \\      
          n \rightarrow p + e^- + {\bar \nu} 
\end{array}$  
\hfill (1.1) 

\

\nnn are not possible ones, we can say that the mirror pictures of the
processes (1.1) are

\ 

\hfill
$\begin{array}{l}
          p_- \rightarrow n_- + e_-^+ + \nu_-        \\          
          n_- \rightarrow p_- + e_-^- + {\bar \nu}_- 
\end{array}$
\hfill (1.2)

\
 
\nnn with the appropriate directions of spins and momenta. The explanation
why process (1.2) have not been observed is that all experiments
have been made with protons $p$ and neutrons $n$ and not with
mirror protons $p_-$ and mirror neutrons $n_-$. Nuclei of Co$^{60}$
contain only protons $p$ and neutrons $n$, and no mirror protons
$p_-$ and mirror neutrons $n_-$, similarly as they contain no
anti-protons and anti-neutrons.

This is the main idea. In the following sections we formulate the
concepts of external and internal improper Lorentz transformations,
especially external and internal space inversion. External
space inversion transforms a particle $a_+$ (at the position
${\bf r}$ with the momentum ${\bf p}$ and spin ${\bf s}$ into
the same particle $a_+$ (with the same internal structure) at
the position $-{\bf r}$ with the momentum $-{\bf p}$ and the
spin ${\bf s}$. Internal space inversion transforms a particle
$a_+$ at the position ${\bf r}$ with the momentum ${\bf p}$
and spin ${\bf s}$ and with the left- (right) handed internal
structure into the particle $a_-$ with the same ${\bf r}$,
${\bf p}$, $\bf s$, but with the opposite, i.e., right- (left)
handed, internal structure. Further, we develop the concept
of total space inversion which transforms a particle $a_+$ with
given ${\bf r}$, ${\bf p}$, $\bf s$ and given internal handedness
into a particle $a_-$ with $- \bf r$, $- \bf p$, $\bf s$, and the
opposite internal handedness.

Next we postulate that each interaction should be invariant under
total space inversion, though it may not be invariant under external
or internal space inversion separately. We then consider the
interaction part of the Hamiltonian for the beta-decay. This
interaction can be modified in such a way that it is invariant
under total space inversion, while it is not invariant under partial
(external or internal) space inversion. The Hamiltonian, so modified,
is still the same as the old one for ordinary particles. Finally we
discuss further possible consequences that follow from the distinction
between external and internal improper Lorentz transformations.
Many new theoretical possibilities are revealed.

\section{Distinction between External, Internal and Total Improper
Lorentz Transformations}

Let us imagine a classical particle, say a stone, with a left-handed
shape. Consider the motion of such an internally asymmetric particle
with a momentum ${\bf p}$. The mirror image of this process is the
motion of the right-shaped stone with the momentum $-{\bf p}$.
If the original {\it left}-shaped stone rotates in a certain screw-sense,
say left sense, around the axis, defined by the initial direction of motion,
then the mirror image of such a process is the motion of the 
{\it right}-shaped stone with the opposite momentum, and spinning in the
opposite, i.e. right-screw, sense. We shall name the transformation
between the two kinds of motion the space reflection or {\it the
total space reflection or inversion}. Because rotations in the 
three-dimensional space do not interest us here, we shall 
ignore the difference between space reflection and space inversion.

Now imagine such a kind of reflection that ignores the internal
asymmetrical structure of the particle. This reflection transforms
the left-shaped particle with a momentum ${\bf p}$ into the 
left-shaped particle with the momentum $-{\bf p}$; if the particle
is spinning in a certain screw-sense, this sense is also reversed
under that transformation. We name it {\it the external space
reflection or inversion}. The definition of the external space
inversion is not an artificial one, since if the left-shaped stone,
spinning in the left screw-sense, moves with a momentum ${\bf p}$
one may always imagine the motion of the same left-shaped stone
spinning in the opposite screw-sense and moving with the opposite
momentum $-{\bf p}$.

The next possibility is to define a transformation which leaves
a particle's translational and rotational motion unchanged, but
inverses its internal structure with a given handedness into the
structure with the opposite handedness. This transformation is 
the space inversion in the particle's internal system of reference
(which is at rest with respect to the particle), which we shall name
{\it the internal space inversion}. It must be clearly stressed here
that the rotation of the particle is with respect to the external frame of
reference, hence the internal transformation does not affect the
particle`s screw-sense rotation.

We shall now use simple symbolism to illustrate the three kinds
of space inversion. By definition, space inversion $P$ is a
transformation that changes a geometrical point ${\bf r} = (x,y,z)$ into
the point $-{\bf r}$. If the point $\rr$ moves with velocity $\bf v$,
its velocity is changed into $- \bf v$ under $P$. Hence
\bear
               P :  && {\bf r} \rightarrow \rr ' = - \rr \nonumber \\
                      && {\bf v} \rightarrow {\bf v }' = - {\bf v} \nonumber
\ear
A geometrical object $A(\rr)$ (a tensor) defined at the position $\rr$
transforms under the space inversion after the well-known rules for
transformation of tensors.

Let us consider a {\it classical} particle which is a physical object,
not a geometrical point. It has a finite size and has both translational
and rotational degrees of freedom. Its state of motion is defined
by coordinates $\rr$ of its mass centre, momentum ${\bf p}$ and 
intrinsic angular momentum $\bf s$ (pure rotation or spinning of
the particle) as functions of time. Under space inversion the three
quantities transform as
$$ \begin{array}{lll}
    P:  & \rr \rightarrow {\rr}' = - \rr      & {\rm vector} \\
       & {\bf p} \rightarrow {\bf p}' = - {\bf p}  & {\rm vector} \\
       & {\bf s} \rightarrow {\bf s}' = {\bf s}   & {\rm pseudovector}
\end{array}$$      
The state of a {\it particle} is, however, not completely characterized by
the quantities $\rr$, $\pp$, and $\ssb$, for the full description we must also
take into account the particle's internal structure. This structure generally
transforms under space inversion, which means that in the internal
coordinate frame each point $a$ with coordinates ${\vec \xi} = (\xi_x ,
\xi_y , \xi_z)$ is transformed into the point $a'$ with coordinates
$-{\vec \xi}$:
\bear 
         P : \quad && a \rightarrow a' = P a \nonumber \\
                     && {\vec \xi} \rightarrow {\vec \xi}' = - {\vec \xi} \nonumber
\ear
In the laboratory coordinate system the same point $a$ has the
coordinates ${\bf r} + {\vec \xi}$. By bearing in mind the upper
transformation performed in the internal frame, we can construct
the following types of transformation performed in the laboratory
system:

\ \ \ (a) total space inversion
\bear
           P_T : \quad && a \rightarrow a'' = P_T a \nonumber \\
               && \rr + {\vec \xi} \rightarrow \rr '' + {\vec \xi}'' =
            - \rr - {\vec \xi} \nonumber
\ear
\ \ \ (b) external space inversion
\bear
           P_E : \quad && a \rightarrow a''' = P_E a \nonumber \\
               && \rr + {\vec \xi} \rightarrow \rr ''' + {\vec \xi}''' =
            - \rr + {\vec \xi} \nonumber
\ear
\ \ \ (c) internal space inversion
\bear
           P_I : \quad && a \rightarrow a' = P_I a \nonumber \\
               && \rr + {\vec \xi} \rightarrow \rr '+ {\vec \xi}' =
            -\rr - {\vec \xi} \nonumber
\ear
The last two transformations are only partial space inversions in
such a sense that either $\rr$ is reversed and $\vec \xi$ left
unchanged (b), or $\rr$ is left unchanged and ${\vec \xi}$ is reversed (c).
Obviously
$$P_T = P_E P_I = P_I P_E$$

Here $\rr$ is the external coordinate, describing the particle's position
in a fixed (laboratory) frame, and $\vec \xi$ is a coordinate associated with
the internal structure of the particle. Since the internal structure
is not determined by the single coordinate $\vec \xi$ but (roughly
speaking) with the distribution of matter within the particle,
we shall use the symbol $\alpha$ in order to denote the parameter
of internal inversion. $\alpha$ has two discrete values; $+1$ and
$-1$. They belong to two opposite states of internal inversion. In
other words, $\alpha$  denotes a set of all internal degrees of freedom
which suffer the internal space inversion. The total state of motion of a
particle with respect to the laboratory system is described by
parameters $\rr$, $\pp$, $\ssb$, $\alpha$. The space inversion yields:

\ 

\hfill
$\begin{array}{ll}
      P_T :  \quad & (\rr , \pp , \ssb , \alpha) \rightarrow 
                 (\rr ' , \pp '  , \ssb ' , \alpha ' ) = 
                 (-\rr , -\pp , \ssb , -\alpha) \\
      P_I :  \quad  & (\rr , \pp , \ssb , \alpha) \rightarrow 
                (\rr ' , \pp '  , \ssb ' , \alpha ' ) = 
                (\rr , \pp , \ssb , -\alpha) \\
      P_E : \quad   & (\rr , \pp , \ssb , \alpha) \rightarrow 
                 (\rr ' , \pp '  , \ssb ' , \alpha ' ) = 
                 (-\rr , -\pp , \ssb , \alpha) 
\end{array}$
\hfill (2.1)

\                  

\nnn By the following example we make the concept of the transformations
$P_T$, $P_E$, $P_I$ as clear as possible. An artist takes several
pieces of optically active substance. Each piece has asymmetric shape. He
then makes a structure that consists of these pieces. Next, he reverses
the positions of all the pieces in the structure and so obtains another
structure which is a partial mirror image of the previous structure;
this transformation is the external space inversion. He then replaces
all the pieces in the last structure with the pieces that have the shape
of the opposite handedness. If he stops his consideration at this point
he will state that the last transformation is the total space inversion.
However, he can go further. If the pieces consist of the optically
active substance with the left screw-sense he can replace them by
pieces with the right screw-sense. Moreover, he may also distinguish the
reflection of the shape from the reflection of the positions of all
the molecules within the piece. He can proceed to a finer and finer
level of structure.

The example above shows that though the concept of pure
geometrical space
inversion (i.e., the inversion of positions of the separate geometrical
points and the appropriate geometrical objects ---tensors---
defined at these points) is clear to us, we must be careful when applying
the space inversion to the real {\it physical object}. It is true that
partial ---for instance the external--- space inversion is in fact not
the real space inversion in the geometrical sense, because, if we
speak about the partial space inversion, we are aware of the further
finer levels of complexity in the observed system. But, dynamically,
whatever kind of physical system we try to transform by space
inversion, at last we necessarily arrive at the level of complexity where
our knowledge about a more detailed structure comes to an end.
At this level we do not distinguish between external and total
space inversion. In every day language we speak about the
space inversion of a physical system if all degrees of freedom that
we are able to control are reversed. The physical system is usually
composed of sub-systems at different levels of complexity.
Such a physical system can be transformed also by partial space
inversion --- external or internal. Partial space inversion is
intimately connected with our knowledge about the levels of
complexity of the system.

Similarly, as in the case of space inversion, we must also distinguish
between total, internal and external time reversal. We denote them
$T_T$, $T_I$ and $T_E$, respectively. They transform the
physical system in the following way:

\ 

\hfill
$\begin{array}{ll}
       T_T  : \quad  & ({\rr} , t, \pp , \ssb , \tau) \rightarrow 
      ({\rr} '  , t', \pp ', \ssb ', \tau') = ({\rr} , -t, -\pp , -\ssb , -\tau) \\
      T_I  : \quad  & ({\rr} , t, \pp , \ssb , \tau) \rightarrow 
     ({\rr} '  , t', \pp ', \ssb ', \tau') = ({\rr} , t, \pp , \ssb , -\tau) \\
     T_E  : \quad & ({\rr} , t, \pp , \ssb , \tau) \rightarrow 
     ({\rr} '  , t', \pp ', \ssb ', \tau') = ({\rr} , -t, -\pp , -\ssb , \tau)
\end{array}$
\hfill (2.2)

\                  

\nnn where $\tau$ if the parameter that characterizes the internal time.
The philosophy is analogous here as in the case of space inversion.
The combination of all $P$ and $T$ types of transformations are
also possible, but the consideration of these possibilities goes
beyond the purpose of the present paper.

The quantum system in the state $\phi (\rr , t, \ssb , \alpha , \tau)$
transforms under the improper Lorentz transformation as
   $$ \phi ({\rr} , t, \ssb , \alpha , \tau) \rightarrow 
      \phi ({\rr} ', t', \ssb ', \alpha' , \tau')$$
where the transformed coordinates $ \rr , t, \ssb , \alpha , \tau$
are replaced according to (2.1) or (2.2). For instance, the
description of an elementary particle by  state vector
$\phi({\rr} , t, \ssb)$ is incomplete. The incompleteness can be
seen just from the fact that for the particle dynamics, as we 
have constructed, discrete Lorentz transformations do not
form a symmetry group for all interactions. The interaction
that governs the beta-decay is not invariant under space
inversion. We shall show later that this non-invariance can be
ascribed to the incomplete description of the system by
the state vector $\phi ({\rr} , t, \ssb )$. By including the internal
degrees of freedom $\alpha$ we can reformulate the dynamics
of the beta-decay so that it becomes invariant under total
space inversion. This reformulation implies the existence
 of mirror particles, i.e., particles with all other quantum
numbers as spin, charge, mass, etc., unchanged, but with
the reversed quantum number $\alpha$---the parameter of
internal inversion. Because we are now inclined to consider
invariance principles as first principles it seems natural
for us to accept the proposal as a serious possibility, which
has to be checked experimentally.

\section{Invariance of  Interactions Under the Total Space Inversion}

The main postulate and the starting point of this article is that
each physical theory must be invariant under the full group of
Lorentz transformations with the improper transformations included.
If observations show that a class of phenomena exists, which
transformed by any Lorentz transformations are not the possible
phenomena, it means that our theory is not invariant under those
transformations. We can adopt two points of view. First, that
our theory is correct and that nature really possesses asymmetry;
and second, that our theory is incomplete, while nature is
symmetric. History teaches us that the second point of view
has always proved to be more convenient. Thus, in the situation
where our theory appears to be non-invariant under any of the
Lorentz transformations, we must admit that we have not yet
obtained complete knowledge about the observed phenomena.
The theory must be completed by including the internal degrees
of freedom of particles involved in the observed phenomena.
While studying the behavior of processes under the space inversion,
the internal handedness of particles must be taken into account in
those cases in which the inversed process would be otherwise
an impossible process.

The state of a system is represented by the state vector 
$\phi(x,\alpha)$, where $\alpha$ is the set of all internal degrees of
freedom which are due to space inversion and $x = (\rr,t)$. We shall
assume that $x$ and $\alpha$ are independent, which gives 
$\phi(x,\alpha) = \phi(x) \varphi(\alpha)$. Parameter $x$ is included
in order to demonstrate that the state vector $\phi(x)$
is completely represented by the field $\phi(x)$ defined over $x$.
The symbol $\phi(x,\alpha)$ denotes the state vector which is not
completely represented by the field $\phi (x)$ and additional degrees
of freedom $\alpha$ have to be taken into account. Parameter
$\ssb$, denoting spin, is omitted. In a given representation, spin
is determined by the type of field (scalar, vector). The transformations
are
  $$P_T \phi(x,\alpha) = \phi'(x',\alpha') = \phi'(-{\rr}, t, -\alpha) =
        \phi'(-\rr , t) \varphi'(-\alpha)$$
 $$P_E \phi(x,\alpha) = \phi'(x',\alpha') = \phi'(-{\rr}, t, \alpha) =
        \phi'(-\rr , t) \varphi'(\alpha)$$
 $$P_I \phi(x,\alpha) = \phi'(x',\alpha') = \phi'({\rr}, t, -\alpha) =
        \phi'(\rr , t) \varphi'(-\alpha)$$

We postulate that the Hamiltonian $H$ is invariant under total space
inversion $P_T$, which means that both operators commute

\ 

\hfill
$[P_T, H] = 0$ 
\hfill (3.1)

\                  

\nnn From $P_T = P_E P_I = P_I P_E$, $P_E \neq 0$, $P_I \neq 0$ and from
$$ [P_T , H] = [P_E P_I , H] = P_E [P_I , H] + [P_E , H] P_I = 0$$
follows either
  $$ [P_E , H] = 0 \qquad {\rm and} \qquad [P_I , H] = 0$$
or
  $$ [P_E , H] \neq 0 \qquad {\rm and} \qquad [P_I , H] \neq 0$$

Equation (3.1) implies that if $\phi(x,\alpha)$ is an eigenstate of $H$
with the energy $E$, then $P_T \phi (x,\alpha)$ is an eigenstate of the same
$H$ with the same energy. Generally $\phi(x,\alpha)$ and $P_T \phi(x,
\alpha)$ are not the same, and there is a degeneracy with respect to
total space inversion.

Operator $P_E$ either commutes with $H$ or does not commute.
In the first case the states $\phi(x,\alpha)$ and $P_E \phi(x,\alpha)$
are both eigenstates of $H$ with the same energy $E$. We do not
observe any degeneracy due to external space inversion, hence
both $\phi(x, \alpha)$ and $P_E \phi (x, \alpha)$ represent the same
state. Following well-known procedure we conclude that the state
$\phi(x, \alpha)$ has the definite parity which is conserved under
the interaction $H$. In the second case, parity is not a good quantum
number and is not conserved. We stress explicitly that parity
relates to the external part $\phi(x)$ of the total state vector
$\phi(x) \varphi (\alpha)$ and the external parity operator $P_E$.

Whether the external parity is conserved or not depends on the kind
of interaction $H$. The strong interaction, for instance, conserves
parity to a high degree of precision, while the weak interaction violates
the external parity. We would like to show that the non-conservation of
external parity does not necessarily imply the non-invariance
of nature under space inversion. The conventional theory of beta-decay
is not invariant under space inversion, and it can be improved to
become invariant under space inversion by including the internal
degrees of freedom. However, this is only a theoretical possibility,
to be proved or disproved by experiment.

The proposed properties of the strong and the weak interactions will now
be described.

\vs{8mm}

\nnn {\it The strong Interaction}

\vs{2mm}

\nnn The strong interaction is invariant under the transformation $P_E$:
  $$ [P_E , H] = 0 $$
From (3.1) it follows that also $[P_I , H] = 0$. As is experimentally
established, there is no degeneracy with respect to the external space
inversion and we have
   $$P_E \phi(\rr ,t) \varphi(\alpha) = \phi'(-\rr , t) \varphi(\alpha) =
\xi_E \phi (\rr , t) \varphi(\alpha)$$
  $$ P_E^2 \phi(\rr ,t) \varphi(\alpha) = P_E \xi_E \phi (\rr ,t) 
  \varphi(\alpha) =
\xi_E^2 \phi(\rr ,t) \varphi(\alpha) = \phi(\rr, t) \varphi(\alpha)$$
therefore
   $$\xi_E = \pm 1.$$
$\xi_E$ is the external intrinsic parity of the state 
$\phi(\rr ,t) \varphi(\alpha)$.
It is a good quantum number. For instance, different particles have
different external intrinsic parities.

With respect to the internal space inversion $P_I$ we have to admit two
possibilities: either a degeneracy or a definite internal parity. The
possibility of the definite internal parity, i.e. the identity of states
$\phi(x)\varphi(\alpha)$ and $P_I \phi(x) \varphi(\alpha)$, is excluded
for nucleons, because we postulate that $[P_T , H] = 0$ for all
interactions. In the case of the weak interaction this would not be
fulfilled because we could not introduce mirror nucleons. Hence
there is a degeneracy: $\phi(x) \varphi(\alpha)$ and $P_I \phi(x) \varphi(\alpha)$ 
represent two different states, both eigenstates of the same Hamiltonian.
If  $\phi(x) \varphi(\alpha)$ is the state of a nucleon $p$ (denoted also as
$p_+$), then $P_I \phi(x) \varphi(\alpha) = \phi(x)\varphi'(-\alpha)$ is the
state of a mirror nucleon $p_-$.

In order to explain why both degenerate states do not occur in the nucleus,
we have to assume that the strong interaction is strong enough to yield
bound states only if it works on particles of the same kind, otherwise it
gives no bound states. The scheme for the two-body strong interaction is
$$\begin{array}{lll}
 \alpha_1 = 1 ,  \quad & \alpha_2 = 1   \quad   
      & \mbox{strong attractive force} \\
\alpha_1 = 1 , \quad  & \alpha_2 = - 1 \quad  
      & \mbox{weak attractive or repulsive force}\\
\alpha_1 = -1 , \quad & \alpha_2 = 1 \quad 
      & \mbox{\rm weak attractive or repulsive force} \\
\alpha_1 = -1 , \quad & \alpha_2 = - 1  \quad & \mbox{strong attractive force}
\end{array}$$
\nnn The natural conclusion, therefor, is if there are nuclei with $p_+$ and
$n_+$ there must also be nuclei with $p_-$ and $n_-$. Are such nuclei in
the ordinary matter on the earth in small concentrations (like isotopes) or in
large concentrations? Possibly mirror nucleons do not constitute ordinary
matter like anti-nucleons, and could be found only under extreme
conditions, for instance at high energies, in cosmic rays or in the other
parts of the Universe. We have not sufficient information from experiment
to decide which possibility holds, if either.

\vs{8mm}

{\it The Weak Interaction}

\vs{2mm}

In the conventional picture of beta-decay  is governed by the Hamiltonian
which is not invariant under space inversion. But as we have shown it is
necessary to distinguish among external, internal and total space inversion.
Each interaction is, after our postulate, invariant under total space
inversion, though not necessarily under partial space inversion. The
Hamiltonain for the beta-decay (Lee \& Yang, 1957) can be modified
in such a way that it is invariant under total space inversion.

\

\hfill
     $H_{\rm int} = {1\oo \sqrt{2}} \sum_i G_i [{\bar \psi}_p (x,\alpha) \Gamma_i
   \psi_n(x,\alpha)][{\bar \psi}_e (x,\alpha) \Gamma_i (1 + \eta
\gamma_5) \psi_{\nu} (x, \alpha)]$
\hfill (3.2)

\ 

\nnn $\psi(x,\alpha)$ is a field defined formally over the external space-time
coordinate $x$ and the internal parameter $\alpha$. Detailed knowledge
of the field is not necessary for the present considerations. We only
require that $x$ and $\alpha$ are independent and that $\psi(x,\alpha)$
is a solution of the Dirac equation. We can the write formally

\

\hfill
$\begin{array}{l}
    \psi(x,\alpha) = \psi(x) \varphi(\alpha) \\
   {\bar \psi}(x,\alpha) = \varphi^{\dagger} (\alpha) {\bar \psi} (x)
\end{array}$
\hfill (3.3)   

\ 

\nnn where $\psi (x)$ is a Dirac field and $\varphi(\alpha)$ a field associated
with the internal state. Parameter $\eta$ in equation (3.2) depends upon
the values of $\alpha$ of individual particles involved in the beta-decay.
  $$ \eta = \biggl\{ 
\begin{array}{ll}  
  \hs{3mm} 1  \quad & {\rm for} \; \; \alpha_p = \alpha_n = 
  \alpha_e = \alpha_\nu = 1\\
    -1 \quad & {\rm for} \; \; \alpha_p = \alpha_n = \alpha_e = 
    \alpha_\nu = -1
\end{array}$$
If one of the $\alpha's$ differs from the others there is no weak
interaction, or it is qualitatively different. For positive $\eta$ the weak
interaction produces the right-handed beta-decay, and for negative
$\eta$ the left-handed beta-decay, where signatures right-left are
chosen arbitrarily. The situation has its classical analog in a gun,
with a spiral trace inside the barrel. If a bullet has a spiral on its
surface corresponding to the spiral inside the barrel, one obtains
the `interaction' which causes the fired bullet to rotate in a certain
screw-sense. If the spiral of the bullet does not fit the gun's
spiral, there is no such screw-sense interaction and the missile
will not leave the barrel.

Inserting (3.3) into (3.2) the Hamiltonian becomes
$$
  H_{\rm int} = \sum_i {{G_i}\oo {\sqrt(2)}} [\varphi_p^{\dagger}
(\alpha) {\bar \psi}_p (x) \Gamma_i \psi_n (x) \varphi_n (\alpha)]
 [\varphi_e^{\dagger} (\alpha) {\bar \psi}_e (x) 
\Gamma_i (1 + \eta \gamma_5) \psi_{\nu} (x) \varphi_\nu (\alpha)]
$$  
This can be written in a more simple form (see for instance
Muirhead, 1968):
\bear
&&H_{int} = {{G_V}\oo \sqrt{2}} [\varphi_p^{\dagger} (\alpha)
{\bar \psi}_p (x) \gamma_{\lambda} \psi_n (x) \varphi_n (\alpha)]
[\varphi_e^{\dagger}(\alpha) {\bar \psi }_e (x) \gamma_{\lambda} (1 +
\eta \gamma_5) \psi_\nu (x) \varphi_\nu (\alpha)]\hs{.8cm} (3.4) \nonumber \\
 &&  \hs{.7cm} + {{G_A}\oo \sqrt{2}} [\varphi_p^{\dagger} (\alpha)
{\bar \psi}_p (x) i \gamma_{\lambda} \gamma_5 \psi_n (x) 
\varphi_n (\alpha)] [\varphi_e^{\dagger} (\alpha) {\bar \psi}_e (x)
i \gamma{\lambda} \gamma_5 (1 + \eta \gamma_5) \psi_\nu (x)
\varphi_\nu (\alpha)] \nonumber
\ear
where $G_V$ and $G_A$ are vector and axial-vector coupling
constants, respectively. The total space inversion gives
$$\psi(x,\alpha) \rightarrow \psi'(x',\alpha') = P_T \psi (x,\alpha) =
P_E \psi (x) P_I \varphi(\alpha) = \biggl\{
\begin{array}{l}
    \pm \gamma_4 \psi (x) P_I \varphi(\alpha)\\
    \pm i \gamma_4 \psi (x) P_I \varphi(\alpha)
\end{array}$$
$${\bar \psi}(x,\alpha) \rightarrow {\bar \psi}' (x',\alpha') =
   \varphi'^{\dagger} (\alpha') \psi'^{\dagger} (x') \gamma_4 =
  \varphi^{\dagger} (\alpha) P_I^{\dagger} \psi^{\dagger} (x)
  P_E^{\dagger} \gamma_4$$
$$ = \biggl\{ 
\begin{array}{l}
       \varphi^{\dagger} (\alpha) P_I^{\dagger} \psi^{\dagger}
          (\pm \gamma_4) \gamma_4 \\
     \varphi^{\dagger} (\alpha) P_I^{\dagger} \psi^{\dagger}
          (\mp i \gamma_4) \gamma_4
\end{array}$$    
Here $P_E$ is identical with the operator $S$ that performs the space
inversion on the Dirac field. It is equal to $\pm \gamma_4$ or $\pm i
\gamma_4$ for the real or the imaginary parity class, respectively.
The following relations are satisfied:

    \ (I) Real parity class

\ 

\hfill
$\begin{array}{c}  
   P_E P_E = 1 , \qquad P_E^{\dagger} P_E = 1 \\
   P_E = P_E^{-1} = \pm \gamma_4 = P_E^{\dagger}
\end{array}$
\hfill  (3.5)

\
    
   (II)  Imaginary parity class

\ 

\hfill
$\begin{array}{c}  
  P_E P_E = -1 , \quad P_E^{\dagger} P_E = 1 \\
  P_E = - P_E^{-1} = \pm i \gamma_4 = - P_E^{\dagger}
\end{array}$
\hfill  (3.6)

\

\nnn We shall assume that operator $P_I$ is also unitary
   $$P_I P_I^{\dagger} = P_I^{\dagger} P_I = 1$$
Hamiltonian (3.4) consists of the terms which are transformed as
\bear
       && V^\lambda = \varphi^\dagger (\alpha) [{\bar \psi} (x) \gamma_\lambda
    \psi (x) ] \varphi(\alpha)  \hs{4cm} \nonumber \\
    && \hs{1cm}
  \rightarrow V'^\lambda   
  = \varphi'\dagger (\alpha') [{\bar \psi}' (x') \gamma_\lambda \psi' (x')] 
  \varphi' (\alpha') \nonumber \\
        &&\hs{2.3cm}= \varphi^\dagger (\alpha) P_I^\dagger [\psi^\dagger (x) P_E^\dagger
    \gamma_4 \gamma_\lambda P_E \psi (x)] P_I \varphi(\alpha) \nonumber \\
   && \hs{2.3cm} = \varphi^\dagger (\alpha) P_I^\dagger [a_{\lambda \alpha} {\bar \psi}(x)
\gamma_\alpha \psi (x)] P_I \varphi(\alpha) \nonumber \\
&& \hs{2.3cm} = \varphi^\dagger (\alpha) [a_{\lambda \alpha} {\bar \psi} (x) \gamma_\alpha 
\psi (x) ] \varphi(\alpha) \nonumber \\
  && \hs{2.3cm} = a_{\lambda \alpha } V^\alpha \nonumber
\ear
where
$$ (a_{\lambda \alpha}) = \pmatrix{
                        -1 & 0 & 0 & 0 \cr
                         0 &-1 & 0 & 0 \cr
                         0 & 0 &-1 & 0 \cr
                         0 & 0 & 0 & 1  \cr} $$
and where relations (3.5) or (3.6)  and

\

\hfill
$\gamma_4 \gamma_\lambda \gamma_4 = 
        a_{\lambda \alpha} \gamma_\alpha$
\hfill   (3.7)

\ 

\nnn have been used. The quantity in the bracket [ ] is that which enters
the conventional Hamiltonian for the beta-decay. Under 
space-time transformations it is transformed as a four-vector. The
whole term $V^\lambda$, including the internal fields $\varphi(\alpha)$,
also transforms as a four-vector. The left and the right part of the
term $V^\lambda$ may belong to different particles. In such a case
we use a more precise symbol $V_{ab}^\lambda$, where
subscripts $a$, $b$ refer to the fields of two different particles.
We must admit that fields belong to different parity classes. In the
following we study the transformation properties of $V$-type terms
under total space inversion.
\bear
            && V^\lambda \equiv V_{r,r}^\lambda 
                       \rightarrow {{V'}}_{r,r}^\lambda 
                       = V'^\lambda \nonumber \\
 && V^\lambda \equiv V_{i,i}^\lambda 
                       \rightarrow {V'}_{i,i}^\lambda 
                       = V'^\lambda \nonumber 
\ear
\hfill
$\begin{array}{l}
V_{r,i}^\lambda \rightarrow {V'}_{r,i}^\lambda = i \varphi_r^\dagger
(\alpha) [a_{\lambda \alpha} {\bar \psi}_r (x) \gamma_\alpha \psi_i (x)]
\varphi_i (\alpha) = i a_{\lambda \alpha} V_{r,i}^\alpha = i {V'}^\lambda \\
\\
V_{i,r}^\lambda \rightarrow {V'}_{i,r}^\lambda = - \varphi_i^\dagger
(\alpha) [a_{\lambda \alpha} {\bar \psi}_i (x) \gamma_\alpha \psi_r (x)]
\varphi_r (\alpha) = - a_{\lambda \alpha} V_{i,r}^\alpha = - i {V'}^\lambda
\end{array}$
\hfill (3.8)

\ 

Indices $r$ and $i$ denote the real and imaginary parity class respectively.
The symbol $V$ without subscript stands for the condition where two
fields belong to the same parity class.

The other term forms that forms the Hamitloniann is
\bear
  && A^\lambda = i \varphi^\dagger [{\bar \psi} (x) \gamma_\lambda
  \gamma_5 \psi (x)] \varphi(\alpha) 
   \rightarrow A'^\lambda  = i
 \varphi'^\dagger (\alpha') [{\bar \psi}' (x') \gamma_\lambda \gamma_5
 \psi' (x') ] \varphi' (\alpha') \hs{2cm} \nonumber \\
     && \hs{3cm}= i \varphi^\dagger (\alpha) P_I^\dagger [\psi^\dagger (x) 
     P_E^\dagger
  \gamma_4 \gamma_\lambda \gamma_5 P_E \psi (x)] P_I \varphi_I (\alpha)
  \nonumber \\
  && \hs{3cm} = i \varphi^\dagger (\alpha)[a_{\lambda \alpha} {\bar \psi} (x)
      \gamma_\alpha P_E^{-1} \gamma_5 P_E \psi (x) ] \varphi(\alpha)
  \nonumber \\
   && \hs{3cm} = - i \varphi^\dagger (\alpha)[a_{\lambda \alpha} {\bar \psi} (x)
\gamma_\alpha \gamma_5 \psi (x)] \varphi(\alpha) \nonumber \\
  && \hs{3cm} = - a_{\lambda \alpha} A^\alpha \nonumber 
\ear
where relations (3.5) or (3.6),\, (3.7) and
   $$ \gamma_4 \gamma_5 \gamma_4 = - \gamma_5$$
have been used. The term $A^\lambda$ transforms as an axial vector.
The mixed quantities transform as

\ 

\hfill
$\begin{array}{c}
    A_{r,r}^\lambda \rightarrow {A'}_{r,r}^\lambda = A'^\lambda \\
  A_{i,i}^\lambda \rightarrow {A'}_{i,i}^\lambda = A'^\lambda \\
   A_{r,i} \rightarrow {A'}_{r,i} = i a_{\lambda \alpha} A^\alpha =
   i A'^\lambda \\
A_{i,r} \rightarrow {A'}_{i,r} = - i a_{\lambda \alpha} A^\alpha =
  - i A'^\lambda
\end{array}$  
\hfill (3.9)

\ 

\nnn The Hamiltonian $H_{\rm int}$ can be rewritten in a compact notation,
$$\hs{1.3cm} H_{\rm int} = {{G_V}\oo \sqrt{2}} \left ( V_{p,n}^\lambda 
V_{e, \nu}^\lambda + {\eta \oo i} V_{p,n}^\lambda 
A_{e, \nu}^\lambda \right ) + {{G_A}\oo \sqrt{2}} \left (
A_{p,n}^\lambda A_{e, \nu}^\lambda  + i \eta 
A_{p,n}^\lambda V_{e, \nu}^\lambda \right ) \hs{1.2cm} (3.10)$$
This Hamiltonian is invariant under total space inversion, i.e.,
  $$H_{\rm int} = H'_{\rm int} = H_{\rm int}$$
if particles are distributed into parity classes in the following way:
$$\begin{array}{llll}
       p  &  n  & e  & \nu \\
\hline       
       r  &  r  &  r &   r \\
       i  &  i  &  i &   i \\
       r  &  r  &  i &   i \\
       i  &  i  &  r &   r \\
       r  &  i  &  i &   r \\
       i  &   r &  r &   i \\
\hline
\end{array}$$       
This can be verified directly from the form of the Hamiltonian (3.10) by
having in mind the relations (3.8),\, (3.9) and the relations
    $$ \eta' = - \eta$$
\bear
  &&{V'}_{p,n}^\lambda {V'}_{e, \nu}^\lambda = 
  (a_{\lambda \alpha} V_{p,n}^\alpha) (a_{\lambda \beta}
  V_{e, \nu}^\beta) = V_{p,n}^\lambda V_{e, \nu}^\lambda \nonumber \\
  &&{V'}_{p,n}^\lambda {A'}_{e, \nu}^\lambda =
        - V_{p,n}^\lambda A_{e, \nu}^\lambda \nonumber \\
  &&{A'}_{p,n}^\lambda  {V'}_{e,\nu}^\lambda =
      - A_{p,n}^\lambda  V_{e,\nu}^\lambda \nonumber \\
  &&{A'}_{p,n}^\lambda  {A'}_{e,\nu}^\lambda =
    A_{p,n}^\lambda  A_{e,\nu}^\lambda \nonumber
\ear    

Thus we have constructed the Hamiltonian for the beta-decay that is
invariant under total space inversion. The old Hamiltonian for the
beta-decay (Lee \& Yang, 1957) is only a special case of this more
general Hamiltonian. If we apply only the external space inversion
$P_E$, then $\eta' = \eta$, and $H_{\rm int}$ obviously changes
its form Our Hamiltonian is for the ordinary particles $P_+$, $n_+$,
$\nu_+$, the same as the conventional one (except for the additional
$\phi^\dagger (\alpha)$ and $\varphi(\alpha)$), because all these
particles have $\alpha = 1$, and therefor $\eta = +1$.
It does not conserve parity.

The main objection against the scheme of Yang and Tiomno
(Yang \& Tiomno, 1950), namely that it fails in the case of beta-decay,
because the weak interaction is not invariant under space inversion,
is thus surmounted. This forgotten scheme explains the
conservation of baryons by assuming that baryons and leptons
belong to two different parity classes and, further, that
all interactions are invariant under space inversion. Then, indeed,
interactions containing terms which create (annihilate) a
baryon and simultaneously annihilate (create) a lepton
would be excluded, because under inversion a factor $i$ would
appear. However, this scheme does not explain the separate
conservation of electron leptons and myon leptons. Perhaps this
additional complication could be explained by taking into account
the internal degrees of freedom.  For our present purpose it is not
necessary to propose any detailed scheme. We wish to show only
new theoretical possibilities, connected with a formal
existence of two parity classes.

We now return to the main subject, the explanation of beta-decay
in terms of a Hamiltonian that is invariant under total space inversion.
We shall no longer distinguish between real and imaginary
parity class, so it is reasonable to assume that all particles
belong to the real class---according to usual practice.

If in the initial state there is a positron $e_+^+$ and neutron $n_+$,
the Hamiltonian described by equation (3.10) produces the
process

\ 

\hfill
  $e_+^+ + n_+ \rightarrow p_+^+ + {\bar \nu}_+ (\uparrow)$
\hfill   (3.11)

\ 

\nnn The symbol $(\uparrow)$ denotes the positive helicity 
($\ssb \cdot \pp >0$) and
the symbol $(\downarrow)$ the negative helicity ($\ssb \cdot \pp <0$). A
process obtained from (3.11) by external space inversion
  $$e_+^+ + n_+ \rightarrow p_+^+ + {\bar \nu}_+ (\downarrow)  $$
does not occur under the action of $H_{\rm int}$ (equation (3.10)).
But according to our hypothesis a process exist which is obtained
from (3.11) by application of total space inversion:
    $$e_-^+ + n_- \rightarrow p_-^+ + {\bar \nu}_- (\downarrow) $$
This last process is governed by the same Hamiltonian (3.10) as
process (3.11).

Let us sum up. The Hamiltonian $H_{\rm int}$ is such that the initial
state $|e_+^+ n_+ \rangle$ develops into the final state $|p_+^+
{\bar \nu}_+ (\uparrow) \rangle $ and the initial state
$|e_-^+ n_- \rangle$ into the final state $|p_-^+
{\bar \nu}_- (\downarrow) \rangle $. The mirror particles $e_-^+$,
$e_-^-$, ${\bar n}_-$, $p_-^+$, $p_-^-$, $\nu_-$, ${\bar \nu}_-$, etc.,
have all the properties of mass, charge, spin, etc., similar to the
corresponding ordinary particles, except that their behavior is
different in the processes with weak interaction, they have the
opposite internal handedness. {\it Their existence is a logical
consequence of the postulate that each physical theory must be
invariant under total space inversion.} The theory of beta-decay,
formulated with the aid of mirror particles, is indeed invariant
under total space inversion. Mention should be made that the
sufficient condition for the invariance is the existence of
mirror nucleons. The existence of mirror leptons is not necessary
for reflection invariance.

\section{Discussion and Conclusion}

We have shown that non-conservation of parity in beta-decay does
not necessarily imply the mirror asymmetry of basic laws of nature.
If we take into account the structure of nucleons we can imagine
that this structure is asymmetric with respect to space inversion and,
because of this fact, the angular distribution of particles at the
beta-decay is also asymmetric. In other words, what is asymmetric
at the beta-decay are initial conditions, while the weak interaction
itself is symmetric. In order to formulate this idea more
precisely we have made a distinction among external, internal and
total space inversion. The space inversion is said to be total if
applied to all degrees of freedom that we are able to control by
experiment, at least in principle. Whilst discussing reflection
invariance or non-invariance of the beta-decay, conventionally we
have in mind the external space-inversion, and not the total one,
because we have not mentioned the inversion of the internal structure
of the nucleon. In the present paper we have shown that by including
the internal structure of particles involved in the beta-decay we can in
principle restore reflection invariance. Of course, our proposal
must be confirmed expermentally. But in any case, the conclusion that
the non-conservation of parity---the external parity--- means that
nature is not invariant under space inversion is wrong. Non-conservation
of external parity means only that nature is not invariant under
partial space inversion, but reveals nothing about invariance or
non-invariance under total space inversion.

In the present work we have postulated that nature is invariant under
total space inversion. This assumption is justified by the fact that
nature has always appeared to be symmetric in its basic laws. On the
other hand, the exact validity of dynamical symmetries has proved many
times in history to be only an extrapolation. For instance, the
conservation of mechanical energy is only an idealization which holds
in the absence of dissipative forces. One could have said in the early
stages of the development of physics that the invariance under a translation
in time is only approximative. However, by including the internal
translational and rotational degrees of freedom (which manifest themselves
in the heat movement of molecules) the conservation od energy is again
restored. But in the next step one can measure precisely the kinetic
energies of molecules and observe that energy is not strictly conserved.
Firstly because of inelastic atomic excitations and, secondly, because of
the equivalence of mass and energy. Next, the conservation of energy
was seriously questioned in the beta-decay before the discovery of
neutrino. At every stage one could question the invariance of
Lagrangian theories under a translation in time, but such has not been
the behavior of physicists. The strong belief in time translation
invariance forced them to correct their theories, that have at last been
confirmed by experiment. In the case of reflection invariance the situation
is similar. Geometrical space inversion is only an idealization, and
how it could be applied as an active transformation on a dynamical system
which has yet unknown structure in the microdomain cannot be imagined.
In the present work we have seen, in a very rough manner, how the theory
of beta-decay can be formulated to be invariant under space inversion.
This can be achieved by formally including in the description the structure
of nucleons.

\vs{8mm}

\centerline{\it Acknowledgement}

\vs{2mm}

I thank Prof.\,Dr.\,J.\,Strnad, and Dr.\,M.\,Vakselj for very useful
discussions.

\vs{8mm}

\centerline{\it References}

\vs{2mm}

\nnn {\small Fraunfelder, H.R. {\it et al.} (1957). {\it Physical Review},
{\bf 106}, 386.

\nnn Lee, T.D. and Yang, C.N. (1957). {\it Physical Review}, {\bf 105}, 1671

\nnn Muirhead, H. (1965)). {\it The Physics of Elementary Particles}.
Pergamon Press Ltd.

\nnn Salam, A. (1957). {\it Nuovo Cimento}, {\bf 5}, 299.

\nnn Yang, C.N. and Tiomno, J. (1950). {\it Physical Review}, {\bf 79}, 498.

\end{document}